\documentclass[aps,nofootinbib,twocolumn,prl,superscriptaddress,amsmath,amssymb,amsfonts,floatfix]{revtex4-2}
\usepackage{amsmath}
\usepackage{amssymb}
\usepackage{amsfonts}
\usepackage[dvips]{graphicx}
\usepackage{epsfig}
\usepackage{tabularx}
\usepackage{color}

\newcommand{\Omegar}{\Omega_\mathrm{R}}
\newcommand{\Omegan}{\Omega_\mathrm{n}}
\newcommand{\Omegam}{\Omega_\mathrm{m}}
\newcommand{\Omegap}{\Omega_\mathrm{p}}
\newcommand{\Kc}{K_\mathrm{C}}
\newcommand{\Kn}{K_\mathrm{n}}
\newcommand{\Km}{K_\mathrm{m}}
\newcommand{\Kp}{K_\mathrm{p}}

\begin{document}

\title{Bound Bogoliubov quasiparticles in photon superfluids}

\author{Marzena Ciszak}
	\affiliation{CNR-Istituto Nazionale di Ottica, Via Sansone 1, I-50019 Sesto Fiorentino (FI), Italy.}
	\author{Francesco Marino}
	\affiliation{CNR-Istituto Nazionale di Ottica, Via Sansone 1, I-50019 Sesto Fiorentino (FI), Italy.}
	\affiliation{INFN, Sezione di Firenze, Via Sansone 1, I-50019 Sesto Fiorentino (FI), Italy}

\date{\today}

\begin{abstract}

Bogoliubov's description of Bose gases relies on the linear dynamics of noninteracting quasiparticles on top of a homogeneous condensate. Here, we theoretically explore the weakly-nonlinear regime of a one-dimensional photon superfluid in which phonon-like elementary excitations interact via their backreaction on the background flow. The generalized dispersion relation extracted from spatiotemporal intensity spectra reveals additional branches that correspond to \textit{bound} Bogoliubov quasiparticles -- phase-locked collective excitations originating from nonresonant harmonic-generation and wave-mixing processes. These mechanisms are inherent to fluctuation dynamics and highlight non-trivial scattering channels other than resonant interactions that could be relevant in the emergence of dissipative and turbulent phenomena in superfluids.

\end{abstract}

\maketitle

The dynamics of quantum-many body systems can be conveniently described in terms of collective excitations, or quasiparticles, and their interactions. At low temperature, when a system is close to its quantum ground state, only few quasiparticles are excited and collisions between them can be neglected. Within this limit, Bogoliubov derived the spectrum of collective excitations in a dilute homogeneous Bose gas \cite{bogo}. The spectrum is linear at low momenta, which is indicative of the collective (phononic) nature of the excitations, while it becomes quadratic at high momenta where the quasiparticles approach the energy of the individual constituents of the gas. A system the collective excitations of which have these spectral properties satisfies the Landau criterion for superfluidity \cite{legget}.

While noninteracting Bogoliubov quasiparticles provide the microscopic framework of superfluidity, dissipation in isolated quantum fluids arises as an effective phenomenon due to quasiparticle interactions \cite{pylak,robertson}. In three-dimensional (3D) Bose gases, resonant processes involving three Bogoliubov quasiparticles, known as Beliaev-Landau scattering \cite{pines,pit,fed}, provide the main channel for the finite lifetime of the excitations \cite{pit,giorgini,konya}. These three-wave interactions result in the generation of quasiparticles with different energy and momentum, while satisfying the Bogoliubov dispersion relation. Due to the lack of a spectral gap and the convexity of the dispersion curve, energy and momentum for these processes are conserved only in two or more spatial dimensions (see e.g. \cite{rege}). In 1D systems, quasiparticles therefore decay only through higher-order scattering \cite{risti} or via interactions with thermal fluctuations \cite{micheli}.

Compared to the above resonant processes \cite{morgan,hechen,ozeri,huang,yong-li}, nonresonant interactions of Bogoliubov modes have received less attention. 
In nonlinear wave theory, nonresonant interactions are known to conserve energy and momentum giving rise to secondary branches in the dispersion relation. Spectral components lying on these branches are commonly referred to as \textit{bound} waves \cite{longuet,fedo}, since they are phase-locked to \textit{free} waves that satisfy the original dispersion relation and create them via (nonresonant) harmonic-generation or nonlinear mixing processes \cite{phillips,zhang}. Most studies on this topic concern wave turbulence theory and related experiments \cite{zak,naz,falcon1,falcon2,mordant}, where bound waves can explain the observed self-similarity and universal scaling of energy spectra \cite{mich,camp}.

Here, we show that similar processes occur in the fluctuation dynamics of a 1D photon superfluid, resulting in the creation of \textit{bound} Bogoliubov quasiparticles.

In such systems the photons propagating in a nonlinear medium can be seen as a gas of Bose particles weakly-interacting via the material nonlinearity \cite{rica1992,carusottoP,carusottorev}. The slowly varying envelope of the optical field plays the role of the complex order parameter (macroscopic wavefunction), and its fluctuations (small ripples on the transverse optical beam) obey the Bogoliubov dispersion relation \cite{chiao}. Important phenomena such as superfluidity and drag-force cancellation \cite{michel}, nucleation of quantized vortices past an obstacle \cite{vocke2016}, nonequilibrium precondensation \cite{santic} and Bogoliubov quasiparticles \cite{vocke2015,glorieaux} have been experimentally observed. Recent experiments also revealed interference effects between Bogoliubov modes \cite{fontaine} and signatures of quantum depletion \cite{piekarski}, phenomena observed so far only in ultracold atomic gases \cite{cheneau,stamper}.

Bogoliubov's theory of non-interacting quasiparticles aptly describes all the observed phenomena, highlighting the profound analogy between nonlinear photonics and quantum gases. However, beyond the Bogoliubov regime, the scattering and decay of elementary excitations serve as microscopic mechanisms underlying dissipative and complex macroscopic dynamics. These processes remains unexplored for photon superfluids.

In this Letter, we investigate nonresonant interactions between collective excitations in a 1D photon superfluid. As a prototype system we consider a model with both local (Kerr) and nonlocal (thermo-optical) nonlinearities that, depending on the parameters, can support either massless or massive Bogoliubov excitations \cite{marino19,ciszak21}. For largely populated Bogoliubov modes, the excitation spectrum shows additional branches corresponding to spontaneously generated \textit{bound} Bogoliubov quasiparticles. For massless excitations, the new spectral components originate from harmonic generation processes.
On the other hand, massive excitations additionally undergo Stokes and anti-Stokes scattering with a global oscillation of the 1D quasicondensate, producing multiple branches separated by the phonon's rest frequency. These processes provide one of the main channels for the spontaneous decay of excitations in 1D superfluids.

We start from the paraxial wave equation describing a monochromatic optical beam in a 1D nonlinear medium \cite{boyd} \begin{equation}
\partial_z \psi = \frac{i}{2} \partial_{xx}^{2} \psi - i \psi \int dx' \Delta n(x-x')\vert \psi(x',z) \vert^{2} \;
\label{eq1}
\end{equation}
where $\psi$ is the slowly varying envelope of the optical field normalized to the peak intensity $\rho_\mathrm{0}$, the convolution integral with kernel $\Delta n$ accounts for the refractive index change relative to the linear refractive index $n_\mathrm{0}$, and the spatial coordinates have been rescaled to the optical wavenumber $k$ \cite{note1}. The dynamics takes place along the spatial direction $x$, orthogonal to propagation direction of the laser beam, while the propagation coordinate $z$ plays the role of a dimensionless time variable, $t$ (see \cite{note2}).

Linearizing Eq. (\ref{eq1}) around a homogeneous background solution and Fourier transforming both in the dimensionless time and the spatial coordinate we obtain $\Omega^2 = \widehat{\Delta n} (K) K^2 + K^4/4$, where $\Omega$ is the angular frequency of the mode, $K$ its wavenumber, and $\widehat{\Delta n}$ is the spatial Fourier transform of $\Delta n$.
\begin{figure}
\vspace{-0.9cm}
\begin{center}
\includegraphics*[width=1.0\columnwidth]{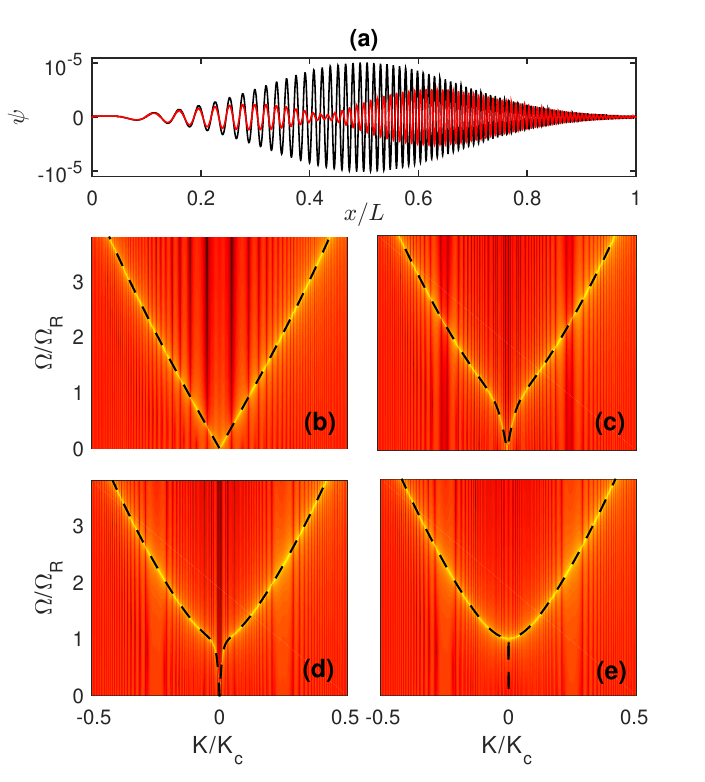}
\end{center} 
\vspace{-0.8cm}
\caption{(a) The initial wavepacket $\psi(x,0)$ constant background subtracted, see text) with $w=10^{-5}$, $K(x)= 3 x /L$ and $\delta=400$ (black trace) and after a time $t=410$ (red trace). 
Panels (b-e) display the numerical dispersion relations reconstructed from the two-dimensional (space-time) Fourier spectrum of the intensity patterns $\vert \psi(x,t)\vert^2$ obtained from Eq. (\ref{eq1}) with $t=z$, $n_2 \rho_\mathrm{0}/n_\mathrm{0} = 10^{-6}$, $\gamma/(n_2 k^2) = 1$ ; (b) $\sigma = 0$; (c) $\sigma= 1$; (d) $\sigma= 10$; (e) $\sigma= 10^{3}$. The dashed lines show the analytical dispersion relation (\ref{disp1}).}
\label{figure1}
\end{figure}

We consider simultaneous local (Kerr) and nonlocal (thermo-optical) nonlinearities $\widehat{\Delta n} = (n_2 + \hat{R}(K))\rho_\mathrm{0}/n_\mathrm{0}$, where $n_2>0$ is the optical Kerr coefficient and $R$ the thermo-optical response function. Optical responses of this kind are found in quantum-dots suspensions \cite{moreels,prb-pbs}, halide perowskites \cite{chen1}, semiconductor materials \cite{torner} and nematic liquid crystals \cite{warenghem}. In atomic superfluids, similar local and nonlocal terms arise in dipolar Bose gases \cite{fattori,vardi}. 

The functional form of $\hat{R}$ depends on the geometry and on the system's boundaries \cite{minovich}. Based on previous theoretical works \cite{conti1,conti2,barad} and experiments \cite{vocke2015,vocke2016,contiexp}, we assume a Lorentzian response $\hat{R} = (\gamma/k^2)\frac{\sigma^2}{1 + \sigma^2 K^2}$, where $\sigma$ is the dimensionless length-scale of the thermo-optical nonlinearity and $\gamma/k^2$ its effective strength.
The dispersion relation now reads
\begin{equation}
\Omega^2 = \Omegar^2 \frac{\sigma^2 K^2}{1+\sigma^2 K^2} + c_\mathrm{s}^2 K^2 \left(1 + \frac{K^2}{\Kc^2} \right) \; ,
\label{disp1} 
\end{equation}
where $\Omegar=\sqrt{\frac{\gamma}{k^2 n_\mathrm{0}} \rho_\mathrm{0}}$ and, in analogy to purely local photon-fluids ($\gamma = 0$), we define the dimensionless sound speed as $c_\mathrm{s} = \sqrt{n_2 \rho_\mathrm{0}/n_\mathrm{0}}$ and critical wavenumber $\Kc= 2 c_\mathrm{s}$, separating the linear and quadratic regime of the dispersion relation (\ref{disp1}).
Since $\Omega^2$ is always positive, the system is neutrally stable to perturbations of all wavenumbers, hence supporting propagating collective excitations ($\Omega^2<0$ would correspond to exponentially-growing modes characteristic of linearly-unstable flows).

\begin{figure*}
\begin{center}
\includegraphics*[width=2.0\columnwidth]{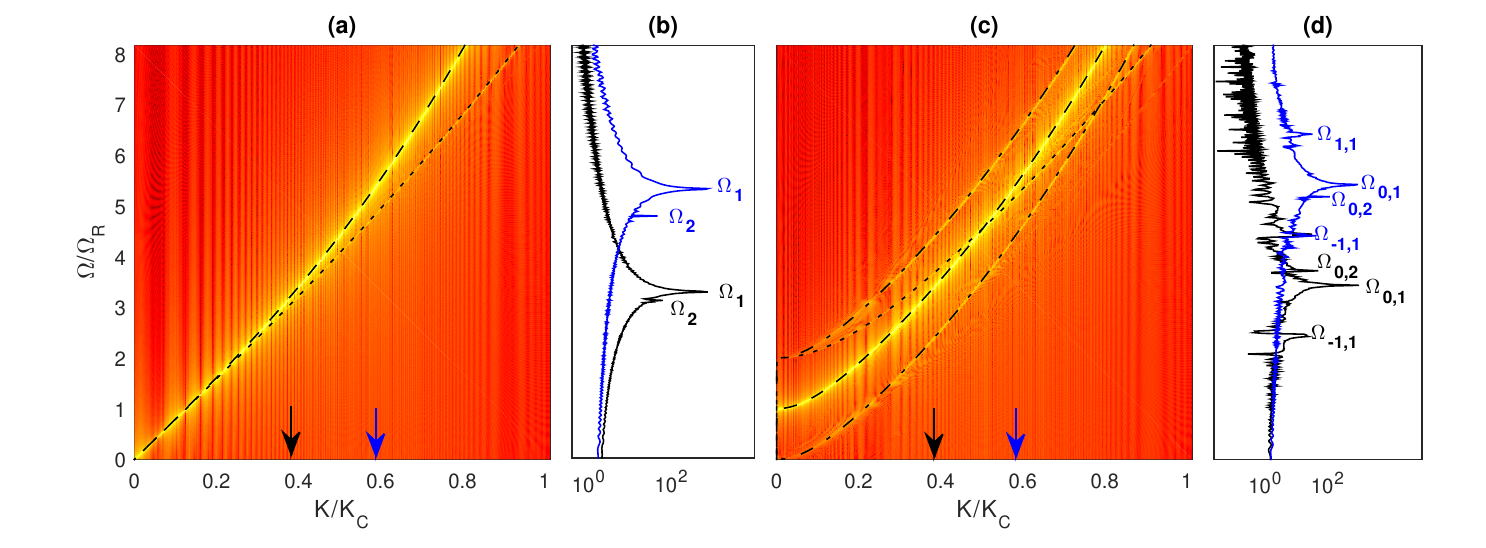}
\end{center} 
\caption{(a) Numerical dispersion relation of the intensity patterns of $\vert \psi(x,t)\vert^2$ with $w=10^{-2}$, $K(x)= 4 x /L$ in the local case $\sigma=0$ and (b) the correspondent frequency Fourier spectra for the wavenumbers $K=0.39 \, \Kc$ and $K = 0.58 \, \Kc$, indicated by the black and blue arrows. Numerical dispersion relation and frequency spectra for $\sigma=10^3$ are shown in (c) and (d). Black lines depict the analytical dispersion relations obtained by (\ref{disp1}) and (\ref{gendisp}) with $\Omega$ and $K$ normalized to $\Omegar$ and $\Kc$, respectively: dashed lines in (a) and (c) correspond to the main branch given by (\ref{disp1}) or, equivalently, by $\Omega_\mathrm{0,1}$ in (\ref{gendisp}). Dotted lines are the second-harmonic branches $\Omega_\mathrm{0,2}$. The dash-dotted curves in panel (c) are the frequency-splitted dispersion branches $\Omega_\mathrm{-1,1}$ and $\Omega_\mathrm{1,1}$. Other parameters as in Fig. \ref{figure1}.}
\label{figure2}
\end{figure*}

For $\gamma = 0$, Eq. (\ref{disp1}) reduces to the Bogoliubov dispersion relation for (massless) collective excitations in a weakly-interacting Bose gas. On the other hand, for $\gamma>0$ the dispersion relation becomes non-convex, and in the limit of $\sigma K \gg 1$ describes massive Bogoliubov quasiparticles with rest frequency $\Omegar$ \cite{marino19}. Such a regime can be reproduced by means of suitable background optical beams comprising wavevectors only of $K \gg 1/\sigma$ (a procedure experimentally implemented in Ref. \cite{danieleSN}) or, similarly, by tuning the ratio between the system's size $L$ and the thermo-optical scale $\sigma$, as we will show in the following.    

To characterize the dispersion relation both in the linear and weakly-nonlinear regime we integrated Eq. (\ref{eq1}) for a numerical time $t_{max} \approx 2.6 \times 10^4$ using a pseudo-spectral second-order Strang splitting method with truncating $2/3$ dealiasing rule. The integration was performed over a spatial domain length $L \approx 1638$ ($N=2^{15}$ grid points with spatial resolution $\Delta x=0.05$) with periodic boundary conditions. 

To set meaningful values for the nonlinear coefficients, we consider a colloidal suspension of PbS nanoparticles in a $C_\mathrm{2} Cl_\mathrm{4}$ solution ($5.9$ nm-size, concentration $6.06 \mathrm{\mu M}$) shined by a laser beam at $\lambda=1.539 \mathrm{\mu m}$. A nonlinear Kerr coefficient $n_2 \approx 4.5 \times 10^{-11} \mathrm{cm^2/W}$ independent of the optical intensity up to $25 \mathrm{MW/cm^2}$ has been measured, together with a linear absorption coefficient $\alpha = 2.5 \mathrm{cm^{-1}}$ and a change in the refractive index with respect to the temperature, $\vert \beta \vert = 0.9 \times 10^{-3} \mathrm{K^{-1}}$ \cite{prb-pbs}. The thermo-optic coefficient $\gamma$ is given by $\gamma=\alpha \vert \beta \vert /\kappa$, where $\kappa$ is the thermal conductivity of the material. Using the thermal conductivity of $C_2 Cl_4$, $\kappa = 0.103 \mathrm{W/m K}$ \cite{handbook} and $n_\mathrm{0}=1.5$, we obtain $\gamma/k^2 \sim 5.9 \times 10^{-11} \mathrm{cm^2/W}$, close to the observed value of $n_2$. The strength of these nonlinearities can be precisely tuned, and even improved up to values of $10^{-7} \mathrm{cm^2/W}$, by changing the type, concentration, and size of the nanoparticles, using a different solvent, or operating at different wavelengths or temperature \cite{moreels,prb-pbs,pbse}. Here, we take $n_2 \rho_\mathrm{0}/n_\mathrm{0} = 10^{-6}$ (a value attainable for average intensities $\rho_0 \sim 22 \mathrm{kW/cm^2}$) and, for simplicity, we set the thermo-optical coefficient equal to the Kerr one, i.e. $\gamma/(n_2 k^2) = 1$.  However, the results we will show below are not critically dependent on these parameters.

The system is initialized with a linearly-chirped, Gaussian wavepacket on top of a spatially-homogeneous background of fixed amplitude $\psi(x,0)= \rho_\mathrm{0}^{1/2}(1 + \varepsilon(x))$ with $\varepsilon(x) = w \, \exp (- i K(x) x) \exp(-x^2/\delta^2)$ (see Fig. \ref{figure1}(a)), which allows us to populate several spatial modes and observe their evolution in a single realization. A noisy initial condition would yield similar effects, though we have verified that the resulting dispersion curves are generally less defined, especially at higher wavenumbers. The dispersion relation is extracted from the two-dimensional (space-time) Fourier spectrum of the intensity patterns $\vert \psi(x,t)\vert^2$. 
\begin{figure}
\begin{center}
\includegraphics*[width=1.0\columnwidth]{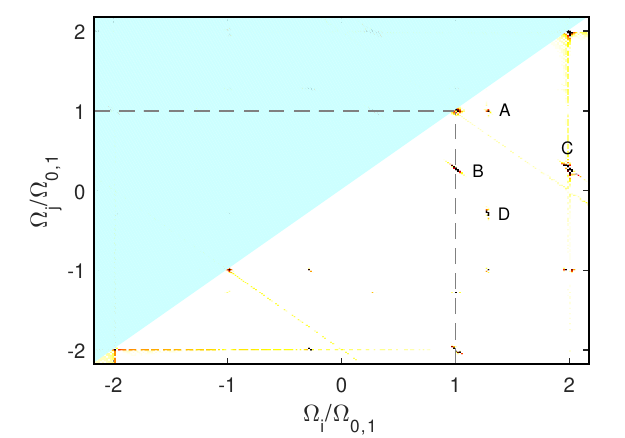}
\end{center} 
\caption{Bicoherence for a monocromatic excitation $\varepsilon(x)= w \, \cos (3 x)$ in the nonlocal case $\gamma/(n_2 k^2) = 1$ and $\sigma=10^3$ for $w=10^{-2}$. All frequencies are normalized to the carrier frequency $\Omega_\mathrm{0,1}$. Since the diagonal is a line of symmetry, $\mathcal{B}(\Omega_i,\Omega_j)$ is plotted only in the half plane $\Omega_i>\Omega_j$ (white-background triangular region).}
\label{figure3}
\end{figure}

The results both in the local and nonlocal cases are shown in Fig. \ref{figure1}(b-e). For $\sigma = 0$ the spectrum is gapless and corresponds to the usual Bogoliubov dispersion relation. For finite $\sigma$ instead the dispersion curves are non-convex and exhibit a dip centered at $K=0$. The dip becomes increasingly sharp as the characteristic length scale of the thermo-optical interaction, $\sigma$, increases. 
When $2 \pi \sigma \gg L$, the condition $\sigma K \gg 1$ holds for all modes supported by the system, in particular for the lower wavenumber $K_{min}=2 \pi/L$, and a gap forms with frequency close to $\Omegar$. 

Increasing the initial population of each mode, secondary branches of collective excitations arise in the dispersion relation (see Fig. \ref{figure2}). All these branches can be interpreted in terms of bound Bogoliubov quasiparticles originating from two different nonlinear processes. The first involves the propagation of higher-harmonics of free Bogoliubov modes satisfying ($\Omega_\mathrm{n}$, $K_\mathrm{n}$)=($n \Omega$,$n K$) with $n=2, \, 3 \,...$. At difference with resonant harmonics, these excitations do not propagate with their own phase velocity, but with the one of the related carrier modes $\Omegan/\Kn=\Omega / K$. We notice that by construction $\Omegan(K)= n \Omega(K/n)$ and therefore all branches are fully determined by the dispersion relation of linear excitations, $\Omega_\mathrm{1}(K) \equiv \Omega(K)$.
An example is illustrated in Fig. \ref{figure2}(a), where we show the numerical dispersion relation corresponding to the local case $\gamma=0$ (massless excitations). Below the main branch $\Omega(K)$, a secondary branch given by $\Omega_\mathrm{2}(K)$ is formed, which corresponds to quasiparticles created via a second-harmonic process. 
At a fixed $K$, we have ($n-1$) peaks of bound quasiparticles in the frequency Fourier spectrum that can be gradually populated depending on the energy injected into the system. Two of these peaks, corresponding to vertical cuts of Fig. \ref{figure2}(a) made at the locations indicated by the arrows, are shown in Fig. \ref{figure2}(b). 

The second mechanism for the formation of bound quasiparticles is the nonresonant mixing between an arbitrary free Bogoliubov excitation and a dominant mode of the system ($\Omegap$, $\Kp$). This interaction results in the emergence of new spectral components at ($\Omegam$, $\Km$)=($\Omega \pm m \Omegap$,$K \pm m \Kp$), with $m=1, \, 2 \,...$. While such a dominant mode does not exist in the local system, the weakly nonlinear regime of Eq. (\ref{eq1}) for $\gamma >0$ is characterized by a global oscillation of the 1D quasicondensate. For sufficiently large $L$, the oscillation frequency is close the rest frequency of the excitations $\Omegar$. Since $\Omegam (K)=\Omega(K \pm m \Kp)\pm m\Omegap$ with $(\Omegap, \Kp) \approx (\Omegar, 0)$, the generalized dispersion relation describing all branches of bound Bogoliubov excitations is
\begin{equation}
\Omega_\mathrm{m,n}(K)=n\Omega\left(\frac{K}{n}\right) \pm m\Omegar \, ,
\label{gendisp} 
 \end{equation}
where $n$ is the harmonic index and $m$ identifies the correspondent frequency-splitted sub-branches. Using this notation the harmonic branches $\Omegan(K)$ are denoted by $\Omega_\mathrm{0,n}(K)$.
First signatures of this structure can be seen in the spectrum in Fig. \ref{figure2}(c) obtained for $\gamma >0$ (massive excitations). Apart from the second-harmonic branch, other secondary branches of bound waves are also visible on each side of the dispersion relation. For each $K$, the interaction between free Bogoliubov excitations and the quasicondensate global mode generates spectral peaks at distances close to $\Omegar$ (see Fig. \ref{figure2}(d)). The process is reminiscent of Stokes and anti-Stokes scattering, where the anti-Stokes sideband on the blue side of the spectrum implies an energy transfer from the quasicondensate mode to Bogoliubov quasiparticles (frequency up-conversion), and vice versa for the Stokes (red) sideband. Similarly to bound quasiparticles associated to higher-harmonics, excitations on these branches are phase-locked to the corresponding free-wave components (in a frame rotating at frequency $m \Omegar$ they propagate with the phase velocity of free Bogoliubov modes). 
The width of the spectral peaks in Fig. \ref{figure2}(b),(d) provides the decay rate of the quasiparticles. We observe values between $1.2-1.7 \times 10^{-4}$ of the order of the frequency spacing between discrete modes $(\partial \Omega/\partial K)K_{min} = 2.8-2.95 \times 10^{-4}$, compatible with the regime of mesoscopic turbulence \cite{pre82-056322}.

\begin{figure}
\vspace{-1cm}
\begin{center}
\includegraphics*[width=1.0\columnwidth]{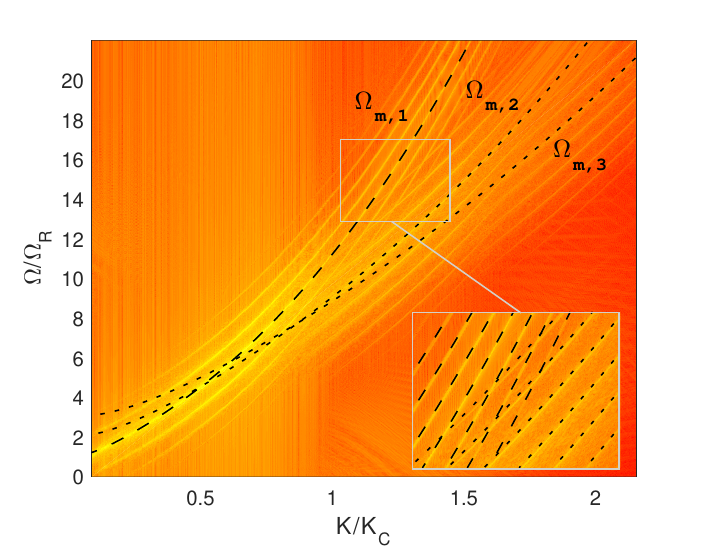}
\end{center} 
\vspace{-.6cm}
\caption{Numerical dispersion relation of the intensity patterns of $\vert \psi(x,t)\vert^2$ in the nonlocal case $\gamma/(n_2 k^2) = 1$ and $\sigma=10^3$ for $w=6 \times 10^{-2}$. The black lines show the analytic curves $\Omega_\mathrm{0, n}$ corresponding to the central branches of each band of harmonics $\Omega_\mathrm{m,n}$: $n=1$ (dashed line), $n=2$ and $n=3$ (dotted lines). The inset show a magnification of a region of the wavenumber-frequency space.}
\label{figure4}
\end{figure}

The phase-coherence between free and bound Bogoliubov components can be detected by computing the normalized third-order correlation function (bicoherence)
\begin{equation}
\mathcal{B}(\Omega_i,\Omega_j) = \frac{\langle \tilde{\psi}(x,\Omega_i)\tilde{\psi}(x,\Omega_j)\tilde{\psi}^*(x,\Omega_i+\Omega_j) \rangle}{ \langle |\tilde{\psi}(x,\Omega_i)\tilde{\psi}(x,\Omega_j)|^2 \rangle\langle |\tilde{\psi}^*(x,\Omega_i+\Omega_j)|^2\rangle}
\end{equation}
where $\tilde{\psi}(x,\Omega)$ denotes the temporal Fourier transform of $\psi$ and $\langle . \rangle$ the averaging over space and time windows of the time-series. The bicoherence quantifies the proportion of quasiparticle energy for any frequency pair ($\Omega_i,\Omega_j$) that is phase coupled to generate a third quasiparticle at energy $\Omega_k$, such that $\Omega_k = \Omega_i + \Omega_j$. The bicoherence for a monochromatic excitation corresponding to a specific Bogoliubov mode (carrier mode) is depicted in Fig. \ref{figure3}. Points on the diagonal $\Omega_i=\Omega_j$ are indicative of phase-coherence of a given signal with itself. The point ($1,1$) corresponds to self-coherence of the carrier mode and implies the generation of a bound excitation via second harmonic process. Self-coherence of such excitation is found at ($2,2$). On the diagonal we also observe the points ($-1,-1$) and ($-2,-2$) that represent the negative-frequency counterparts of the above modes. Note that $\mathcal{B}(\Omega_i,\Omega_j)$ is symmetric about the diagonal and we can thus focus only on half of the frequency plane (white-background region in Fig. \ref{figure3}). The point ($2,-1$) is indicative of a second-harmonic bound quasiparticle corresponding to the process $\Omega_\mathrm{0,2}-\Omega_\mathrm{0,1} = \Omega_\mathrm{0,1}$. Similarly, the point $A$=($\Omega_\mathrm{1,1}/\Omega_\mathrm{0,1}$, $1$) signals the phase-coherence between the carrier Bogoliubov mode and the anti-Stokes sideband splitted at about $\Omegar$ with respect to the carrier frequency. The corresponding process $\Omega_\mathrm{1,1}-\Omega_\mathrm{0,1} = \Omegar$ is also highlighted by the existing phase-coherence between the global mode at frequency $\Omegar$ and the carrier $\Omega_{0,1}$, shown by the points $B$ and $D$. The bicoherence reveals several other points, some of which represent the negative frequency counterparts of the scattering processes above described, as well as higher-order processes. For instance, point $C$ signifies phase coupling between the global mode $\Omegar$ and the bound excitation at frequency $\Omega_\mathrm{0,2}$, resulting in the generation of a new quasiparticle at $\Omega_\mathrm{1,2} = \Omega_\mathrm{0,2} + \Omegar$.

For largely populated Bogoliubov modes, the excitations energy is distributed over several new branches in the wavenumber-frequency space. An example is shown in Fig. \ref{figure4} for the case of massive excitations. While each branch is exactly described by Eq. (\ref{gendisp}) with the corresponding indices and without the need for free parameters (see Inset), the space-time Fourier spectrum presents a high degree of complexity, characterized by intersecting bands of bound excitations. Stokes and anti-Stokes branches of order $m$ appear as organized in bands each one corresponding to a given harmonic number $n$. In Fig. \ref{figure4} one can distinguish three harmonic bands (first, second and third harmonic), each one consisting of $7$ Stokes and anti-Stokes branches with $m=-3,-2...3$.  Such a complicated structure highlights energy transfer channels in photon superfluids other than resonant interactions that, for the nonlocal case, can occur for any wavenumber even in one dimension. 

In conclusion, we studied the weakly-nonlinear regime of a photon superfluid, in which nonresonant phonon interactions give rise to bound Bogoliubov quasiparticles. It is important to highlight that bound quasiparticles are the manifestation of the backreaction of collective excitations on the underlying superfluid. The excitations perturb the initially homogeneous flow, and this modulated flow subsequently alters their propagation, giving rise to an effective interaction mechanism. The resulting bound excitations, which are revealed by a structure of extra branches in the dispersion relation, could be detected and charaterized by existing experiments in both paraxial Kerr \cite{fontaine,piekarski} and polaritonic fluids of light \cite{claude2023}, or in quantum nonlinear optical setups \cite{Peyronel}, where they could emerge at the few-photon level. Beyond the optical domain, a compelling direction lies in the exploration of two-component atomic superfluids, where the coexistence of massless and massive phonon excitations has been experimentally observed \cite{ferrari22}. Bound quasiparticles evidence channels other than resonant interactions for the emergence of dissipation and complex dynamics in 1D Bose gases. Recent experiments with surface gravity waves have shown that bound waves could explain the observed scaling of turbulent energy spectra \cite{mich,camp}. Bound quasiparticles could act similarly in the context of Bogoliubov wave turbulence, where the backreaction on the underlying condensate, at the origin of these secondary excitations, determines the dynamics \cite{tsubota15,zhu}.


\begin{thebibliography}{99}
\bibitem{note1} While a quantum treatment is not necessary to explain these phenomena, the problem could be addressed within the quantum description of paraxial light propagation in purely local Kerr media, developed in \cite{pra92}.
\bibitem{note2} The physical propagation coordinate $z_\mathrm{ph}=z/k$ can be interpreted as an effective time variable $\tau=(n_\mathrm{0}/c)z_\mathrm{ph}$, where $c$ is speed of light in vacuum (see, e.g. \cite{marino19}). Accordingly, $t=\omega \tau$, where $\omega=c k/n_0$ represents the angular frequency of the laser beam.
\end{thebibliography}
\end{document}